\documentclass[11pt]{article} 

\usepackage{latexsym}
\usepackage{amsmath}
\usepackage{amssymb}
\usepackage{epsfig}
\usepackage{epic}
\usepackage{eepic}
\usepackage{colordvi}

\newcommand{\rem}[1]{}
\newcommand{\ignore}[1]{}
\newcommand{\comment}[1]{}

\def\maps11{\stackrel {1-1}{\longmapsto}}

\newlength{\Iwidth}
\newlength{\Zwidth}

\newcommand{\qed}{%
        \hfill
	\setlength{\fboxsep}{0pt}%
	\setlength{\fboxrule}{1pt}%
	\framebox{{\rule{5pt}{6pt}}}%
}

\comment{

\newtheorem{THEOREM}{Theorem} 
\newenvironment{theorem}{\begin{THEOREM} \hspace{-.85em} {\bf :} \rm}%
                        {\end{THEOREM}} 
\newtheorem{THEOREM2}[THEOREM]{Theorem} 
                        {\end{THEOREM2}} 
\newtheorem{COROLLARY}[THEOREM]{Corollary} 
\newenvironment{corollary}{\begin{COROLLARY} \hspace{-.85em} {\bf :} \rm}%
                          {\end{COROLLARY}} 
\newtheorem{LEMMA}{Lemma} 
\newenvironment{lemma}{\begin{LEMMA} \hspace{-.85em} {\bf :} \rm}%
                      {\end{LEMMA}} 
\newtheorem{LEMMA2}[LEMMA]{Lemma}
                      {\end{LEMMA2}} 
\newtheorem{CLAIM}{Claim} 
\newenvironment{claim}{\begin{CLAIM} \hspace{-.85em} {\bf :} \rm}%
                      {\end{CLAIM}} 
\newtheorem{PROPOSITION}{Proposition} 
\newenvironment{proposition}{\begin{PROPOSITION} \hspace{-.85em} {\bf :} \rm}%
                      {\end{PROPOSITION}} 
\newtheorem{EXAMPLE}{Example} 
\newenvironment{example}{\begin{EXAMPLE} \hspace{-.85em} {\bf :} \rm}%
                        {\end{EXAMPLE}} 
\newenvironment{proof}{\noindent {\bf Proof:} \hspace{.677em}}%
                      {\qed} 
                      {} 
                      {} 
                      {}
\newtheorem{DEFINITION}{Definition} 
\newenvironment{definition}{\begin{DEFINITION} \hspace{-.85em} {\bf :} \rm}%
                           {\end{DEFINITION}} 
\newtheorem{DEFINITION2}[DEFINITION]{Definition}
                           {\end{DEFINITION2}} 
\newtheorem{OBSERVATION}{Observation} 
\newenvironment{observation}{\begin{OBSERVATION} \hspace{-.85em} {\bf :} \rm}%
                           {\end{OBSERVATION}} 

\newtheorem{REMARK}{Remark} 
\newenvironment{remark}{\begin{REMARK} \hspace{-.85em} {\bf :} \rm}%
                       {\end{REMARK}} 
\newtheorem{EXERCISE}{Exercise} 
                         {\end{EXERCISE}}

}

\usepackage{ifthen}

\newtheorem{theorem}{Theorem}[section]

\newtheorem{lemma}{Lemma}

\newtheorem{claim}[theorem]{Claim}

\newtheorem{corollary}[theorem]{Corollary}
\newtheorem{observation}[theorem]{Observation}
\newtheorem{xmpl}{Example}

\newcounter{spuriouscounter}

\newtheorem{lemma*}[spuriouscounter]{Lemma}
\newtheorem{corollary*}[spuriouscounter]{Corollary}

\newenvironment{proof}{\noindent {\bfseries Proof.} }{\hfill\qed\vspace{1em}}
               {\hfill\qed\vspace{1em}}
               {\hfill\qed\vspace{1em}}
               {\hfill\qed\vspace{1em}}

\newsavebox{\saveboxtxt}

\newsavebox{\progbox}
\newsavebox{\progtitlebox}
\newlength{\titleskip}
\newlength{\codeindent}
\newlength{\linenumwidth}
\newlength{\pauseskip}
\newlength{\indentskip}

\newcounter{lynecount}
\newcounter{startflag}
\newcommand{\defaultlinenumtext}{10.\hspace{12pt}}
\newenvironment{program}[2][\defaultlinenumtext]
  {\addvspace{1em}\begin{lrbox}{\progtitlebox}{\bfseries #2}\end{lrbox}
   \setcounter{lynecount}{0}
   \setcounter{startflag}{0}
   \setlength{\pauseskip}{8pt}
   \setlength{\indentskip}{1cm}
   \setlength{\codeindent}{8mm}
   \setlength{\titleskip}{1em}
   \settowidth{\linenumwidth}{#1}
  }
  {\ifthenelse{\value{startflag} = 0}{\startprog}{}
    \end{tabbing}\end{minipage}\end{lrbox}
    \begin{center}\fbox{\usebox{\progbox}}\end{center}
    \vspace{1em}
  }

\newcommand{\startprog}
   {\setcounter{startflag}{1}
    \begin{lrbox}{\progbox}
    \begin{minipage}[b]{\textwidth}\begin{tabbing}
    \hspace{\codeindent}\=%
    \hspace{\linenumwidth}\=\hspace{\indentskip}\=%
    \hspace{\indentskip}\=\hspace{\indentskip}\=%
    \hspace{\indentskip}\=\hspace{\indentskip}\=\kill
    \usebox{\progtitlebox}\rule[-\titleskip]{0pt}{0pt}\\
   }
\newcommand{\lyne}[1][\}]
  {\ifthenelse{\value{startflag} = 0}{\startprog}{\rule{\codeindent}{0pt}\\}
   \refstepcounter{lynecount}
   \ifthenelse{\equal{#1}{\}}}{}{\label{#1}}
   \>\thelynecount .\>
  }
   {\end{minipage}\end{lrbox}\usebox{\saveboxtxt}}

\newcommand{\negativeEdges}[1]{}
\newcommand{\remove}[1]{}
\newcommand{\fullversion}[1]{}

\begin{document}
\bibliographystyle{plain}

\author{ Shlomo Moran\thanks{
Computer Science dept., Technion, Haifa
32000, Israel. {\tt moran@cs.technion.ac.il}} \and Sagi Snir\thanks{
Department of Mathematics, University of California,
  Berkeley, CA 94720, USA {\tt ssagi@math.berkeley.edu}}}

\title{ Efficient Approximation of Convex Recolorings
 \footnote{A preliminary version of the results in this paper
appeared in \cite{mrtc-nov03}.}
}

\maketitle

\begin{abstract}

A coloring of a
 tree is convex if the
vertices that pertain to any color induce a connected subtree; a
partial coloring (which assigns colors to some of the vertices) is
convex if it can be completed to a convex (total) coloring. Convex
coloring of trees arise in areas such as phylogenetics,
linguistics, etc. eg, a perfect phylogenetic tree is one in which
the states of each character induce a convex coloring of the tree.
Research on perfect phylogeny is usually focused on finding a tree
so that few predetermined partial colorings of its vertices are
convex.

When a coloring of a tree is not convex, it is desirable to know
"how far" it is from a convex one.  In ~\cite{MS04}, a
 natural measure for this distance, called {\em the recoloring
   distance} was defined: the minimal number of color changes at the
vertices needed to make the coloring convex. This can be viewed as
minimizing the number of ``exceptional vertices" w.r.t. to a
closest convex coloring. The problem was proved to be NP-hard even for
colored string.\\
In this paper we continue the work of \cite{MS04}, and
present 
a 2-approximation algorithm of convex recoloring of
strings whose running time $O(cn)$,
where $c$ is
the number of colors and $n$ is the size of the input,
and an $O(cn^2)$-time
3-approximation algorithm for convex recoloring of trees.
\end{abstract}

\eject
\setcounter{page}{1}

\newcommand{\XX}{{\cal X}}
\newcommand{\CC}{{\cal C}^*}
\newcommand{\CCv}{{\cal C}_v^*}
\newcommand{\CCvi}{{\cal C}_{v_i}^*}
\newcommand{\DD}{{\cal D}}
\newcommand{\EE}{{\cal E}}
\newcommand{\st}{{\hspace{1pt}:\hspace{1pt} }}
\newcommand{\opth}{\widehat{opt}}
\newcommand{\Rh}{\widehat{R}}
\newcommand{\PPk}{{\cal PART}_k}

\section{Introduction }
A phylogenetic tree is a tree which represents the course of
evolution for a given set of species. The leaves of the tree are
labelled with the given species. Internal vertices correspond to
hypothesized, extinct species.
A {\em character} is a biological attribute shared among all the
species under consideration, although every species may exhibit a
different {\em character state}.  Mathematically, if $X$ is the set of
species under consideration, a character on $X$ is a function $C$ from
$X$ into a set $\cal C$ of character states. A character on a set of
species can be viewed as a {\em coloring} of the species, where each
color represents one of the character's states.  A natural biological
constraint is that the reconstructed phylogeny have the property that
each of the characters could have evolved without reverse or
convergent transitions: In a reverse transition some species regains a
character state of some old ancestor whilst its direct ancestor has
lost this state. A convergent transition occurs if two species possess
the same character state, while their least common ancestor possesses
a different state.
%

In graph theoretic terms, the lack of reverse and convergent
transitions means that the character is {\em
 convex} on the tree:
for each state of this character, all
species (extant and extinct) possessing that state
induce a  single {\em block}, which is a maximal monochromatic
subtree.
%
Thus, the above discussion implies that in a
phylogenetic tree, each character is likely to be convex or "almost
convex".
This make convexity a fundamental property in the context of
phylogenetic trees to which a lot of research has been dedicated
throughout the years. The {\em Perfect Phylogeny }  (PP) problem, whose
complexity was extensively studied (e.g.
\cite{Gusfield91,KanWar:SICOMP95,IJFCS:AFB1,KannanWarnow97,
ICALP:BodlaenderFW1992, NPsteel}),  seeks
for a phylogenetic tree  that is simultaneously
convex on each of the input characters.
 {\em
Maximum parsimony}  (MP) \cite{Fit,san75} is a very popular tree
reconstruction method that seeks for a tree which
minimizes the parsimony score defined as the number of mutated edges
summed over all characters
(therefore, PP is a special case of MP). \cite{GGPSW96} introduce
another criterion to estimate the distance of a phylogeny from
convexity. They define the {\em phylogenetic number }
as the maximum number of connected components a single
state induces on the given phylogeny
(obviously, phylogenetic number one corresponds to a perfect
phylogeny).
Convexity is a desired property in other areas of classification,
beside phylogenetics.
For instance,
in \cite{NATURE2000, BFY:RECOMB2001} a method called {\em TNoM} is used
to classify genes, based on data from gene expression extracted
from  two types of  tumor tissues.
The method finds a
separator on a binary vector, which minimizes the number of ``1'' in
one side and ``0'' in the other, and thus defines a convex vector of minimum
Hamming distance to the given binary vector.
In ~\cite{ATDFS:PNAS2004}, distance from convexity is used
(although not explicitly)
to show strong connection between strains of
Tuberculosis and their human carriers.


In a previous work ~\cite{MS04}, we defined
and studied a natural distance from a
given coloring to a convex one: the {\em recoloring distance}. In the
simplest, unweighted model, this distance is 
the minimum number of color changes at the vertices needed to
make the given coloring convex (for strings this reduces to
Hamming distance from a closest convex coloring). This model was
extended to a weighted model, where changing the
color of a vertex $v$ costs a nonnegative weight $w(v)$. 
The most general model studied in \cite{MS04} is
the {\em non-uniform} model, where the cost of coloring vertex $v$ by a
color $d$ is an arbitrary nonnegative number $cost(v,d)$. 

It was shown in \cite{MS04}
that finding the recoloring distance in the unweighted model is
NP-hard even for strings (trees with two leaves), and few
dynamic programming algorithms for exact solutions of 
few variants of the problem were presented.

In this work we present two polynomial time, constant ratio
approximation algorithms, one for strings and one for
trees. Both algorithms are for the weighted (uniform) model.
The algorithm for strings is based on a lower bound technique which
assigns penalties to colored trees. The penalties can be computed in
$O(cn)$ time, and once a penalty is computed, a
recoloring whose cost is smaller
than the penalty is computed in linear time.
The 2-approximation follows by
showing that for a string, the penalty is at most twice the cost of an
optimal convex recoloring.
This last result does not hold for trees, where a different technique
is used.
The algorithm for trees is based on a
recursive construction that uses a variant of the local
ratio technique \cite{Reuven,BE85}, 
which allows adjustments of the underlying tree topology
during the recursive process. 

The rest of the paper is organized as follows.
In the next section we
present the notations and define the models used.
In Section \ref{sec:lower} we define the notion of penalty which
provides
lower bounds on the optimal cost of convex recoloring of any tree. 
In Section ~\ref{sec:apx-str}, we present the 
2-approximation algorithm for 
the string. In Section ~\ref{sec:apx-tree} we briefly
explain the local ratio
technique, 
and present the 3-approximation algorithm
for the tree.
We conclude and point out future research directions in
Section~\ref{sec:conc}.

\section{Preliminaries}
\label{sec:prelim}
A colored tree is a pair $(T,C)$ where $T=(V,E)$ is a tree with vertex
set $V=\{v_1,\ldots,v_n\}$, and $C$ is a {\em coloring} of $T$, i.e. -
a function from $V$ onto a set of colors ${\cal C}$.
For a set $U \subseteq V$,
$C|_U$ denotes the restriction of $C$ to the vertices of $U$, and
$C(U)$ denotes the set $\{C(u):u\in U\}$. For a subtree
$T'=(V(T'),E(T'))$
of $T$, $C(T')$ denotes the set $C(V(T'))$.  A {\em block}
in a colored tree is a
maximal set of vertices which induces a monochromatic subtree.  A {\em
$d$-block} is a block of color $d$.  The number of $d$-blocks is
denoted by $n_b(C,d)$, or $n_b(d)$ when $C$ is clear from the context.
A coloring $C$ is said to be {\em convex} if $n_b(C,d)=1$ for every
color $d\in {\cal C}$.  The number of $d$-{\em violations} in the
coloring $C$ is $n_b(C,d)-1$, and the total number of {\em violations}
of $C$ is $\sum_{c\in{\cal C}} (n_b(C,d)-1)$.  Thus a coloring $C$ is
convex iff the total number of violations of $C$ is zero  
(in \cite{SICOMP:FBL03}
the above sum, taken over all characters, is used as a
measure of the distance of a given phylogenetic tree from perfect
phylogeny).
%

The definition of convex coloring is extended to {\em partially
colored} trees, in which
the coloring $C$ assigns
colors to some subset of vertices $U \subseteq V$, which is
denoted by $Domain(C)$. A partial coloring
is said to be convex if it can be extended to a total
convex coloring (see \cite{SSbook03}).
Convexity  of partial and total coloring have simple characterization
by the concept of {\em carriers}: For a subset $U$ of $V$,
$carrier(U)$ is the minimal subtree that contains $U$. For a colored
tree $(T,C)$ and a color $d\in C$, $carrier_T(C,d)$ (or $carrier(C,d)$
when $T$ is clear) is the carrier of $C^{-1}(d)$.
\remove{ if $d\in C(V)$, then $carrier(C,d)$ is a {\em carrier of
$C$}. }
We say that $C$ has the {\em disjointness property} if
for each pair of colors $ \{d,d'\}$
 it holds that $carrier(C,d)\cap carrier(C,d')=\emptyset$.
It is easy to see that a total or partial coloring
$C$ is convex iff it has the disjointness property
(in \cite{DreSte92} convexity is actually defined by the
disjointness property).

When some (total or partial) input coloring $(C,T)$ is given,
any other coloring $C'$ of $T$ is viewed as
a {\em recoloring} of the input coloring $C$.
We say that a recoloring $C'$ of $C$ {\em retains} (the color of) a
vertex $v$ if $C(v)=C'(v)$, otherwise $C'$ {\em overwrites} $v$. 
Specifically, a recoloring $C'$ of $C$ overwrites a vertex $v$ either by
changing the color of $v$, or just by {\em uncoloring} $v$.
We say that
$C'$ retains (overwrites)
a set of verices $U$ if it retains (overwrites resp.)
every vertex in $U$.
For a recoloring $C'$ of an input coloring $C$, $\XX_C(C')$
(or just $\XX(C')$) is the set of
the vertices overwritten by $C'$, i.e.
$$\XX_C(C')=\{v\in V:\left [ v\in Domain(C)\right ]\bigwedge
\left [ (v\notin Domain(C')~) \vee (C(v)\neq C'(v)~)\right ]\}.$$

With each recoloring $C'$ of $C$ we associate a {\em cost}, denoted as
$cost_C(C')$ (or $cost(C')$ when $C$ is understood), which is
the number of vertices overwritten by $C'$, i.e.
$cost_C(C')=|\XX_C(C')|$. 
A coloring  $C^*$ is an {\em optimal convex recoloring of $C$}, or in
short an
{\em optimal recoloring of $C$},
 and $cost_C(C^*)$ is
 denoted by $OPT(T,C)$,
if $C^*$ is a convex coloring of $T$, and
$cost_C(C^*)\leq cost_C(C')$ for any other convex coloring $C'$ of
$T$.

The above cost function naturally generalizes to 
the {\em weighted} version:
the input is a triplet
$(T,C,w)$, where $w:V\rightarrow {\mathbb R}^+\cup\{0\}$ is a
weight function
which assigns to each
vertex $v$ a nonnegative weight $w(v)$.
For a set of vertices $X$,
$w(X)=\sum_{v\in X}w(v)$. 
The cost of a convex recoloring $C'$ of $C$ is
$cost_C(C')=w(\XX(C'))$, and $C'$ is an optimal convex recoloring if it
minimizes this cost.\\
The above unweighted and weighted cost models are {\em uniform}, in the
sense that the cost of a recoloring is determined by the set of
overwritten vertices, regardless the specific colors involved.
\cite{MS04} defines also a
more subtle {\em non uniform model}, which
is not studied in this paper.

Let $AL$ be an algorithm which receives as an input a weighted
colored tree
$(T,C,w)$ and outputs a convex recoloring of $(T,C,w)$,
and let $AL(T,C,w)$
be the cost of the convex recoloring output by $AL$.
We say that $AL$ is an {\em $r$-approximation} algorithm for
the convex tree recoloring problem if for all inputs $(T,C,w)$ it holds
that $AL(T,C,w)/OPT(T,C,w)\leq r$ \cite{GarJoh79,Hoc97}.
\ignore{447
An optimization problem
is {\em fully p-approximable} \cite{PazMoran81},
or has a {\em fully polynomial
time approximation scheme} \cite{GarJoh79},
if for each $\varepsilon$ there is an algorithm which
provides an $\varepsilon$-approximation to it in time which is
polynomial in $n$ and $\frac{1}{\varepsilon}$.
We note that  by the the NP completeness proof in ~\cite{MS04}, the
problem is 
 not fully p-approximable unless $P\ne NP$:
This follows by the
fact that the value of the optimal solution is bounded by the input
size, using an observation in \cite{PazMoran81}.
endignore 447}

We complete this section with a definition and
a simple observation which will be useful
in the sequel.
Let $(T,C)$ be a colored tree.
A coloring $C^*$ is an {\em expanding}
recoloring of $C$ if in each block of $C^*$ at least one vertex
$v$ is retained
(i.e., $C(v)=C^*(v)$).

\begin{observation}
\label{obs:expanding}
let $(T=(V,E),C,w)$ be a weighted colored tree, where $w(V)>0$.
Then there exists an
expanding optimal convex recoloring of $C$.
\end{observation}
\begin{proof}
Let $C'$ be an optimal recoloring of $C$ which uses a minimum
number of colors (i.e. $|C'(V)|$ is minimized). We shall
prove that $C'$ is an expanding recoloring of $C$.

Since $w(V)>0$, the claim is trivial if $C'$ uses just one
color.  So assume for contradiction
 that $C'$ uses at
least two colors, and that for some color $d$ used by $C'$, there
is no vertex $v$ s.t. $C(v)=C'(v)=d$.
Then there must be an edge $(u,v)$ such that $C'(u)=d$
but $C'(v)=d'\neq d$.
Therefore, in the uniform cost model,
the coloring $C''$ which is identical to
$C'$ except that all vertices colored $d$ are now colored by $d'$ is
an optimal recoloring of $C$
which uses a smaller number of colors - a contradiction.
\end{proof}

In view of Observation \ref{obs:expanding} above, we assume in the
sequel (sometimes implicitly) that
the given optimal convex recolorings are expanding.

\section{Lower Bounds via Penalties}\label{sec:lower}
In this section we present a
general lower bound on the recoloring distance of weighted 
colored trees.
Although for a general tree this bound can be fairly
poor,
in the next section we show that for strings it is at least half the
optimal cost, and then we use this fact to
obtain a 2-approximation algorithm for strings.

Let $(T,C,w)$ be a weighted colored tree.
For a color $d$ and $U \subseteq V(T)$  let:
\[
penalty_{C,d}(U)=w(U\cap \overline{C^{-1}(d)})+
w(\overline{U}\cap{C^{-1}(d)})
\]
Informally, when the vertices in $U$ induce a subtree,
$penalty_{C,d}(U)$
is the total weight of the vertices which must be
overwritten to make $U$ the unique $d$-block in the coloring:
a vertex $v$ must be overwritten
either if $v\in U$ and $C(v)\neq d$, or if $v\notin U$ and
$C(v)=d$.

\remove{
\BlueViolet      { BlueViolet  Approximate PANTONE 2755}
\Periwinkle      { Periwinkle  Approximate PANTONE 2715}
\CadetBlue       { CadetBlue  Approximate PANTONE (534+535)/2}
\CornflowerBlue  { CornflowerBlue  Approximate PANTONE 292}
\MidnightBlue    { MidnightBlue  Approximate PANTONE 302}
\NavyBlue       { NavyBlue  Approximate PANTONE 293}
\RoyalBlue       { RoyalBlue  No PANTONE match}
\Blue            { Blue  Approximate PANTONE BLUE-072}
\Cerulean        { Cerulean  Approximate PANTONE 3005}
\Cyan            { Cyan  Approximate PANTONE PROCESS-CYAN}
\ProcessBlue     { ProcessBlue  Approximate PANTONE PROCESS-BLUE}
\SkyBlue        { SkyBlue  Approximate PANTONE 2985}
\Turquoise       { Turquoise  Approximate PANTONE (312+313)/2}
\TealBlue        { TealBlue  Approximate PANTONE 3145}
\Aquamarine      { Aquamarine  Approximate PANTONE 3135}
\BlueGreen       { BlueGreen  Approximate PANTONE 320}
endremove}

\begin{figure}
\begin{center}
\input{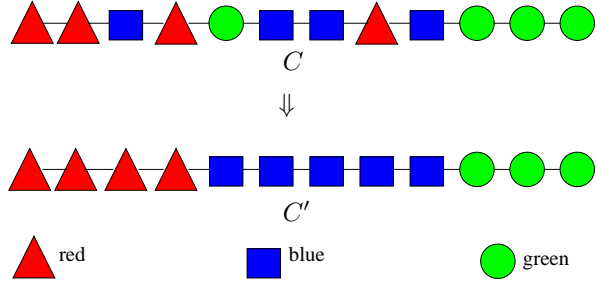}
\end{center}
\caption{$C'$ is a convex recoloring for $C$ which defines the
following penalties: $p_{\Green{green}}(C')=1$,
 $p_{\Red {red}}(C')=2$,
$p_{\Blue {blue}}(C')=3$
\label{fig:penalty}}
\end{figure}

The penalty of a given convex recoloring is sums of the
penalties of every colored block:
Let $C'$ be a convex
recoloring of $C$. Then:
$$penalty_C(C')= \sum_{d \in {\cal C}}penalty_{C,d}(C^{'-1}(d))$$
Figure~\ref{fig:penalty} depicts the calculation of a penalty
associated with a convex recoloring $C'$ of $C$.

In the sequel we assume that the input colored tree
$(T,C)$ is fixed, and omit it
from the notations.

\begin{claim}\label{cla:2cost_C(C')}
$penalty(C')=2cost(C')$
\end{claim}
\begin{proof}
From the definitions we have
\begin{eqnarray*}
penalty(C')&=&\sum_{d\in{\cal C}} w\left (
\{v\in V:C'(v)=d \text{ and }C(v)\neq d\}\cup
\{v\in V:C'(v)\neq d \text{ and }C(v)= d\}\right ) \\
&=&2w(\{v\in V:C'(v)\neq C(v) \})=2cost(C')
\end{eqnarray*}
\end{proof}

As can be seen in Figure~\ref{fig:penalty}, $penalty(C')=6$ while
$cost(C')=3$.\\

For each color $d$, $p^*_d$ is the penalty 
of a block which minimizes the penalty for $d$:

$$ p^*_{d}= \min \{penalty_{d}(V(T')): T' \mbox{ is a 
subtree of } T\}$$

\begin{corollary}
\label{cor:penalty}
For any recoloring $C^*$ of $C$,
$$
\sum_{d\in {\cal C}} p^*_d \leq
\sum_{d\in{\cal C}} penalty_d(C')=2cost(C').
$$
\end{corollary}
\begin{proof}
The inequality follows from the definition of
$p^*_d$, and the equality from Claim \ref{cla:2cost_C(C')}.
\end{proof}

\begin{figure}[t]
\begin{center}
\fbox{
\epsfig{file=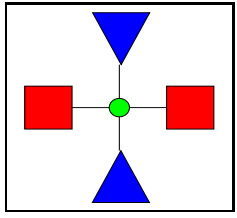,width=1in,height=1in,angle=0}
} 
\end{center}
\caption{At least one of the lateral big blocks (triangle or
  rectangle) needs to be recolored.
\label{fig:poor-bound}}
\end{figure}

Corollary \ref{cor:penalty} above provides a lower bound on the cost
of convex recoloring of trees. It can be shown that this lower bound
can be quite poor for trees, that is: $OPT(T,C)$ can
be considerably larger than $(\sum_{d\in{\cal C}} p^*_d)/2$. For
example, any convex recoloring of the tree in
Figure~\ref{fig:poor-bound}, will recolor at least one of the big
lateral blocks in the tree, while $(\sum_{d\in{\cal C}}
p^*_d)/2$ in
that tree is the weight of the (small) central vertex (the circle).
However in the next section we show that this bound
can be used to obtain a polynomial time 2-approximation for
convex recoloring of
strings.

\section{A $2$-Approximation Algorithm for Strings}\label{sec:apx-str}
Let a weighted colored string $(S,C,w)$, where
$S=(v_1,\ldots,v_n)$, be given.
For $1\leq i\leq j\leq n$, $S[i,j]$ is
the substring $(v_i,v_{i+1},\ldots,v_j)$ of $S$.
The algorithm starts by finding for each $d$ a
substring $B_d=S[i_d,j_d]$ for which $penalty_d(S[i_d,j_d])=p^*_d$.
It is not hard to verify
that $B_d$ consists of a subsequence of consecutive vertices
in which the difference between the total weight of
$d$-vertices and the total weight of other vertices
(i.e. $w(B_d\cap C^{-1}(d))~-~w(B_d\setminus C^{-1}(d))$) 
is maximized, 
and thus $B_d$
can be found in linear time.
We say that a vertex $v$ is {\em covered by color} $d$ if it belongs
to $B_d$. $v$ is {\em covered} if it is covered by some
color $d$, and it is {\em free} otherwise.
\newcommand{\HC}{{\hat C}}

We describe below a linear time
algorithm which, given the blocks $B_d$,
defines a convex coloring $\HC$
so that $cost(\HC) < \sum_d p^*_d$, which by
Corollary \ref{cor:penalty}
is a 2-approximation to a minimal convex recoloring of $C$.

$\HC$ is constructed by performing one scan of
$S$ from left to right. The scan consists of at most
$c$ stages, where stage $j$ defines the $j-th$ block of $\HC$, to
be denoted $F_j$, and its color, $d_j$, as
follows.

Let $d_1$ be the color of
the leftmost covered vertex
(note that $v_1$ is either free or covered by ${d_1}$).
$d_1$ is taken to be the color of the 
first (leftmost)
block of $\HC$, $F_1$, and
$\HC(v_1)$ is set to
$d_1$. For $i>1$, $\HC (v_i)$ is determined as follows: Let
$\hat{C}(v_{i-1})= d_j$. Then if $v_i\in B_{d_j}$ or $v_i$ is free,
then $\hat{C}(v_i)$ is also set to $d_j$. Else, $v_i$ must be a
covered vertex.
Let $d_{j+1}$ be one of the
colors that cover $v_i$.
$\hat{C}(v_i)$ is set to $d_{j+1}$ (and $v_i$ is the first vertex in
$F_{j+1}$).

\begin{figure}[t]
\begin{center}
\fbox{
\input{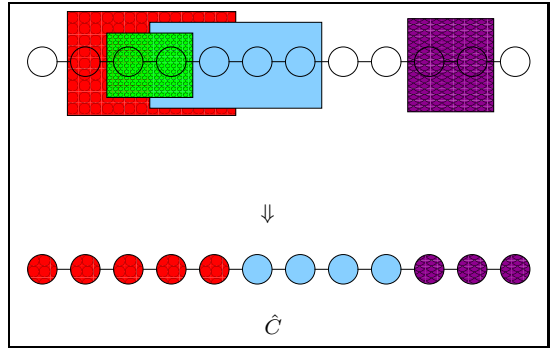}
} 
\end{center}
\caption{ The upper part of the figure shows the optimal blocks on the
string and the lower part shows the coloring returned by the algorithm.
\label{fig:2-approx-coloring}}
\end{figure}

\begin{observation}
$\HC$ is a convex coloring of $S$.
\end{observation}
\begin{proof}
  Let $d_j$ be the color of the $j-th$ block of $\HC$,
  $F_j$, as described above.
  The convexity of $\HC$ follows from the the following
  invariant, which is easily proved by induction: For all $j\geq 1$,
  $\cup_{k=1}^j F_k\supseteq \cup_{k=1}^j B_{d_k}$. This means
  that, for all $j$, no vertex to the right of $F_j$
  is covered by $d_j$, and hence
  no such vertex is colored by $d_j$.
  The observation follows. 
\end{proof}

 Thus it remains to prove
\begin{lemma}
$cost(\HC) < \sum_{d\in {\cal C}} p^*_d$.
\end{lemma}
\begin{proof}
Let $v_i$ be a vertex which contributes to $cost(\hat{C})$.
Then $C(v_i)=d$ and $\HC (v_i)=d'$ for
some  distinct $d', d$.
By the algorithm, either $v_i\in B_{d'}$, or $v_i$
is free. In the first case $v_i$ contributes to both $p^*_d$ and
$p^*_{d'}$, and in the 2nd it contributes to $p^*_d$.
The inequality is strict since
in each block $F_j$ there is at
least one vertex for which the former case holds.
\end{proof}

\section{A 3-Approximation Algorithm for Tree}\label{sec:apx-tree}
\newcommand{\aprx}{$3$-tree-APPROX}
\newcommand{\tbr}{{\bar{T}}}
\newcommand{\xpbr}{{\bar{X'}}}
\newcommand{\tbrz}{{\bar{T_0}}}
\newcommand{\xpbrz}{{\bar{X'}_0}}
\newcommand{\ddz}{d_0}
\newcommand{\rdz}{{r_{d_0}}}
\newcommand{\tht}{\hat{T}}
\newcommand{\xpht}{\hat{X'}}
\newcommand{\vvz}{v_0}
\newcommand{\chig}{C_{high}}
\newcommand{\xhig}{\XX(C_{high})}
\newcommand{\cmed}{C_{medium}}
\newcommand{\xmed}{\XX(C_{medium})}
\newcommand{\cmin}{C_{min}}
\newcommand{\xmin}{\XX(C_{min})}
In this section we  present a polynomial time algorithm which
approximates the
minimal convex coloring of a weighted
tree by factor three. The 
input is a triplet $(T,C,w)$, where
$w$ is a nonnegative weight function and $C$ is a (possibly partial)
coloring whose domain is the set
$support(w)=\{v\in V:w(v)>0\}$.

We firat introduce the notion of covers w.r.t. colored trees.
A set of vertices $X$ is a
{\em convex cover} (or just a cover)
for a colored tree
$(T,C)$ if the (partial) coloring $C_X=C|_{[V\setminus X]}$ is
convex (i.e., $C$ can be transformed to a convex coloring by
overwriting the vertices in $X$).
Thus, if $C'$ is a convex
recoloring of $(T,C)$, then $\XX_C(C')$, the set of vertices
overwritten by $C'$, is a cover for $(T,C)$.
Moreover, deciding whether a subset $X\subseteq V$ is a cover for
$(T,C)$, and constructing a total convex
recoloring $C'$ of $C$ such that $\XX(C')\subseteq X$ in case it is,
can be done in $O(n\cdot n_c)$
time. Also,
 the cost of a recoloring $C'$
is $w(\XX(C'))$. 
Therefore, finding an optimal
convex total recoloring of $C$
is polynomially equivalent to finding
an optimal cover $X$, or
equivalently a partial convex recoloring $C'$ of $C$
so that $w(\XX(C')) =w(X)$ is minimized.

Our approximation algorithm makes use of the local ratio
technique, which is useful for approximating
optimization covering problems such as vertex cover, dominating set,
minimum spanning tree, feedback vertex set and more
\cite{BE85,BBF95,Reuven}.
 We hereafter
describe it briefly:\\
The input to the
problem is a triplet $(V,\Sigma\subseteq 2^V,
w:V\rightarrow \mathbb R^+)$, and the goal is to find a subset
$X\in \Sigma$ such that $w(X)$ is minimized, i.e.
$w(X)=OPT(V,\Sigma,w)=\min\limits_{Y\in \Sigma}w(Y)$ (in our
context $V$ is the set of
vertices, and $\Sigma$ is the set of covers).
The local
ratio principle is based on the following observation (see e.g.
\cite{Reuven}):
\begin{observation}
\label{obs:reuven}
For every two weight functions $w_1,w_2$:
 $$OPT(V,\Sigma,w_1)+OPT(V,\Sigma,w_2) \leq OPT(V,\Sigma,w_1+w_2)$$
\end{observation}
Now, given our initial weight function $w$, we select $w_1,w_2$
s.t. $w_1+w_2=w$ and
$|supprt(w_1)|<|support(w)|$. We first apply the algorithm to find an
$r$-approximation to
$(V,\Sigma,w_1)$ (in particular, if $V\setminus support(w_1)$ is a cover,
then it is an optimal cover to $(V,\Sigma,w_1)$).
Let $X$ be the solution returned for $(V,\Sigma,w_1)$, and assume that
$w_1(X)\leq r\cdot OPT(V,\Sigma,w_1)$. If we could also guarantee that
$w_2(X)\leq r\cdot OPT(V,\Sigma,w_2)$ then by Observation
\ref{obs:reuven} we are guaranteed that $X$ is also an
$r$-approximation for $(V,\Sigma,w_1+w_2=w)$.
The original property, introduced in \cite{BE85}, which was used
to guarantee that $w_2(X)\leq r\cdot OPT(V,\Sigma,w_2)$
is that $w_2$ is {\em $r$-effective}, that is:
for every $X\in\Sigma$
it holds that $w_2(X)\leq r\cdot OPT(V,\Sigma,w_2)$
(note that if $V\in\Sigma$, the above is equivalent to requiring that
 $w_2(V)\leq r\cdot OPT(V,\Sigma,w_2)$).
\begin{theorem}\cite{BE85}
\label{the:localratio}
Given $X\in\Sigma$ s.t. $w_1(X) \leq r\cdot OPT(V,\Sigma,w_1)$.
If $w_2$ is
$r$-effective, then $w(X)=w_1(X)+w_2(X) \leq r \cdot OPT(V,\Sigma,w)$.
\end{theorem}

%
We start by presenting two applications
of Theorem \ref{the:localratio} to obtain a
$3$-approximation algorithm for convex recooloring of strings and a
$4$-approximation algorithm for convex recoloring of trees.

\begin{center}
\fbox{
\begin{minipage}{0.95\textwidth}
$3$-string-APPROX:\\
Given an instance to convex weighted string problem $(S,C,w)$:
\begin{enumerate}
\item If $V\setminus support(w)$ is a cover
then $X\leftarrow V\smallsetminus support(w)$. Else:
\item\label{con:string-violation} Find 3 vertices
$x,y,z\in support(w)$ s.t. $C(x)=C(z)\ne C(y)$ and $y$ lies between
$x$ and $z$.
\begin{enumerate}
\item $\varepsilon \leftarrow \min\{w(x),w(y),w(z)\}$
\item $ w_2(v) = \left\{ \begin{array}{ll}
                        \varepsilon  & \mbox{if } v\in \{x,y,z\} \\
                        0       &\  \mbox{otherwise.}
                      \end{array}
               \right. $
\item $w_1\leftarrow w-w_2$
\item $X\leftarrow 3$-string-APPROX$(S,C|_{support(w_1)},w_1)$
\end{enumerate}
\end{enumerate}

\end{minipage}
}
\end{center}

Note that if a (partial) coloring of a string is not convex
then the condition in \ref{con:string-violation} must hold.
It is also easy to see that $w_2$ is $3$-effective,
since any cover
$Y$ must contain at least one vertex from any
triplet described in condition~\ref{con:string-violation}, hence
$w_2(Y)\geq \varepsilon$ while
$w_2(V)=3\varepsilon$.

The above algorithm cannot serve for approximating convex tree
coloring since in a tree
the condition in \ref{con:string-violation} might not hold 
even if $V\setminus support(w)$ is not a cover.
In the following algorithm we
generalize this condition to one which must hold in any non-convex
coloring of a tree, in the price of increasing the approximation ratio
from 3 to 4.
\begin{center}
\fbox{
\begin{minipage}{0.95\textwidth}
$4$-tree-APPROX:\\
Given an instance to convex weighted tree problem $(T,C,w)$:
\begin{enumerate}
\item If $V\smallsetminus support(w)$ is a cover
then $X\leftarrow V\smallsetminus support(w)$. Else:
\item\label{con:tree-violation} Find two pairs of (not
necessarily distinct) vertices
$(x_1,x_2)$ and $(y_1,y_2)$ in $support(w)$ s.t. $C(x_1)=C(x_2)\ne
(y_1)=C(y_2)$, and $carrier(\{x_1,x_2\})\cap
carrier(\{y_1,y_2\})\neq\emptyset$:
\begin{enumerate}
\item $\varepsilon \leftarrow \min\{w(x_i),w(y_i)\}$, $i=\{1,2\}$
\item $ w_2(v) = \left\{ \begin{array}{ll}
                        \varepsilon  & \mbox{if } v\in \{x_1,x_2,y_1,y_2\} \\
                        0       &\  \mbox{otherwise.}
                      \end{array}
               \right. $
\item $w_1\leftarrow w-w_2$
\item $X\leftarrow 4$-tree-APPROX$(S,C|_{support(w_1)},w_1)$
\end{enumerate}
\end{enumerate}
\end{minipage}
}
\end{center}

The algorithm is correct since if there are no two pairs as described
in step
\ref{con:tree-violation}, then $V\setminus support(w)$ is a cover.
Also, it is easy to see that
$w_2$ is 4-effective. Hence the above
algorithm returns a cover with weight at most $4\cdot OPT(T,C,w)$.

We now describe algorithm
\aprx. Informally, the algorithm 
uses an iterative method, in the spirit of the local ratio technique,
which
approximates the solution of the input $(T,C,w)$ by reducing it to
$(T',C',w_1)$ where $|support(w_1)|<|support(w)|$. Depending on the
given input, this reduction is either of the
local ratio type (via an appropriate
$3$-effective weight function) or, 
the input graph is replaced
by a smaller one which preserves the optimal solutions.

\begin{center}
\fbox{
\begin{minipage}{16cm}
\aprx $(T,C,w)$\\
On
input $(T,C,w)$ of a weighted colored tree, do the following:
\begin{enumerate}
\item If $V\setminus support(w)$ is a cover
then $X\leftarrow V\smallsetminus support(w)$. Else:
\item
\label{ite:reduce}
$(T',C',w_1)\leftarrow REDUCE(T,C,w)$. $\setminus$The function $REDUCE$
guarantees that
$|support(w_1)|<|support(w)|$
\begin{enumerate}
\item $X'\leftarrow$ \aprx$(T',C',w_1)$.
\item
\label{ite:update}
$X\leftarrow UPDATE(X',T)$. $\setminus$The function $UPDATE$ guarantees
that if $X'$ is a 3-approximation to $(T',C',w_1)$, then $X$ is a
3-approximation to $(T,C,w)$.
\end{enumerate}
\end{enumerate}
\end{minipage}
}
\end{center}

Next we describe the functions $REDUCE$ and $UPDATE$,
by considering few cases. In the first two cases
we employ the local ratio
technique.

\noindent
{\bf Case 1:}
$support(w)$ contains three vertices $x,y,z$
such that $y$ lies on the path from $x$ to $z$ and
$C(x)=C(z)\neq C(y)$.\\
In this case we use the same reduction of $3$-string-APPROX:
Let $\varepsilon=\min\{w(x),w(y),w(z)\}>0$. Then
$REDUCE(T,C,w)=(T,C|_{support(w_1)},w_1)$, where
$w_1(v)=w(v)$ if $v\notin\{x,y,z\}$, else
$w_1(v)=w(v)-\varepsilon$.
The same arguments which implies the correctness of $3$-string-APPROX
implies that
if $X'$ is a 3-approximation
for $(T',C',w_1)$, then it is
also a 3-approximation for
$(T,C,w)$,
thus we set $UPDATE(X',T)=X'$.

\begin{figure}[t]
\begin{center}
\fbox{
\epsfig{file=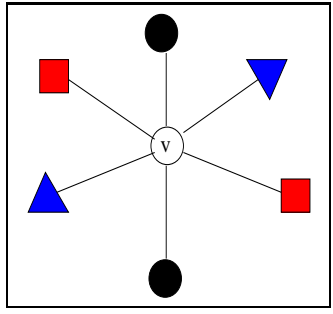,width=1.5in,height=1.5in,angle=0} 
} 
\end{center}
\caption{Case 2: a vertex $v$ is contained in 3 different carriers.
\label{fig:3-approx-case-2}}
\end{figure}

\noindent
{\bf Case 2:}
Not Case 1, and
$T$ contains a vertex $v$ such that $v\in\cap_{i=1}^3 carrier(d_i,C)$
for three distinct colors $d_1,d_2$ and $d_3$ (see Figure
\ref{fig:3-approx-case-2}).\\
In this case we must have that $w(v)=0$ (else Case 1 would hold),
and there
are three {\em designated pairs}
of vertices $\{x_1,x_2\},\{y_1,y_2\}$ and $\{z_1,z_2\}$
such that $C(x_i)=d_1$, $C(y_i)=d_2, C(z_i)=d_3 (i=1,2)$,
and $v$ lies on
each of the three paths connecting these three
pairs (see Figure \ref{fig:3-approx-case-2}).
We set $REDUCE(T,C,w)=(T,C|_{support(w_1)},w_1)$,
where $w_1$ is
defined as follows.\\
Let $\varepsilon=\min\{w(x_i),w(y_i),w(z_i):i=1,2\}$.
Then $w_1(v)=w(v)$ if $v$ is not in one of the designated
pairs, else $w_1(v)=w(v)-\varepsilon$.
Finally, any cover for $(T,C)$
must contain at least two vertices from the set
$\{x_i,y_i,z_i:i=1,2\}$, hence $w-w_1=w_2$ is 3-effective,
and by the local ratio theorem we can set
$UPDATE(X',T)=X'$.

\begin{figure}[bht]
\begin{center}
\input{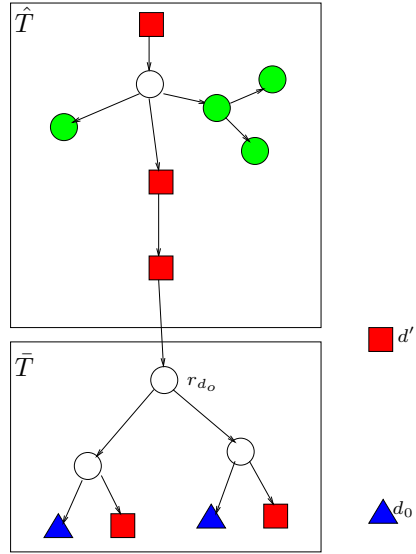}
\end{center}
\caption{Case 3: Not case 1 nor 2. $\bar T$ is the subtree rooted
at $\rdz$ and $\tht=T\setminus \tbr$.
\label{fig:3-approx-case-3}}
\end{figure}

\noindent
{\bf Case 3:}
Not Cases 1 and 2.\\
Root $T$ at some vertex $r$
and for each color $d$
let $r_d$ be the root of the subtree $carrier(d,C)$.
Let $\ddz$ be a color for which the root
$\rdz$ is farthest from $r$. Let
$\tbr$ be the subtree
of $T$ rooted at $r_{d_0}$, and
let $\tht= T\setminus\tbr$ (see Figure
\ref{fig:3-approx-case-3}).
By the definition of $\rdz$, no vertex in $\tht$ is colored by
$\ddz$, and since Case 2 does not hold, there is
 a color $d'$ so that
 $\{\ddz\}\subseteq C(V(\tbr))\subseteq\{\ddz,d'\}$.

\begin{figure}[thb]
\begin{center}
\input{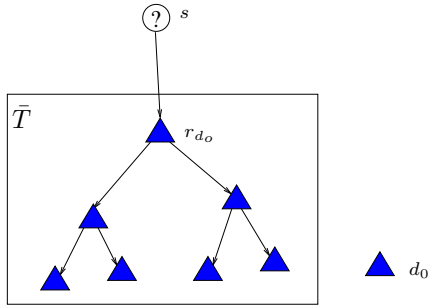}
\end{center}
\caption{Case 3a:
No vertices of $\tht$ are colored by $d'$.
\label{fig:3-approx-case-3a}}
\end{figure}

\noindent
{\bf Subcase 3a:}
$C(V(\tbr))=\{\ddz\}$ (see Figure
\ref{fig:3-approx-case-3a}).\\
In this case, $carrier(\ddz,C)\cap
carrier(d,C)=\emptyset$ for each color $d\neq \ddz$,
and for each optimal solution $X$ it holds that
$ X\cap V(\tbr)=\emptyset$. We set
$REDUCE(T,C,w)\leftarrow (\tht, C|_{V(\tht)},w|_{V(\tht)})$. The
3-approximation $X'$ to $(T',C',w_1)$ is also a 3-approximation to
$(X,C,w)$, thus $UPDATE(X',T)=X'$.

\noindent

We are left with the last case.\\
{\bf Subcase 3b:}
$\rdz\in carrier(\ddz,C)\cap carrier(d',C)$.
See Figure~\ref{fig:3-approx-case-3b}.\\
Observe that
in this case we have $w(\rdz)=0$ and $|support(w)\cap V(\tbr)|\geq 3$,
since $V(\tbr)$ must contain at least
two
vertices colored $\ddz$ and at least one vertex
colored $d'$.
Figure~\ref{fig:3-approx-case-3b} illustrates this case.

\begin{figure}
\begin{center}
\input{3-approx-case-3b_f.pstex_t}
\end{center}
\caption{Case 3b: $\rdz \in T_{\ddz}\cap carrier(d')$
\label{fig:3-approx-case-3b}}
\end{figure}


\begin{observation}
\label{obs:wlog}
There is an optimal convex coloring $C'$ which satisfies the
following: $C'(v)\neq \ddz$ for any $v\in V(\tht)$, and
$C'(v)\in \{\ddz,d'\}$ for any $v\in V(\tbr)$.
\end{observation}
\begin{proof} 
Let $\HC$ be an expanding optimal convex recoloring of $(T,C)$.
We will show that there is
an optimal coloring $C'$ satisfying the lemma such that $cost(C')\leq
cost(\HC)$.
Since $\HC$ is expanding and optimal, at least one vertex in $\tbr$ is
colored either by $\ddz$ or by $d'$. Let $U$ be a set of vertices in
$\tbr$ so that $carrier(U)$ is a maximal subtree all of whose vertices
are colored by colors not in $\{\ddz,d'\}$. Then
$carrier(U)$ must have a neighbor $u$ in $\tbr$ s.t.
$\HC(u)\in\{\ddz,d'\}$. Change the colors of the vertices in $U$ to
$\HC(u)$. This procedure can be repeated until all the
vertices of $\tbr$ are colored by $\ddz$ or by $d'$, without
increasing the cost of the recoloring. A similar procedure can be used
to change the color of all the verticed in $\tht$ to be different from
$\ddz$.
It is easy to see that 
the resulting coloring $C'$ is convex and $cost(C')\leq cost(\HC)$.
\ignore{1108
conditions in the observation. Now, if $\HC$ violates both
conditions, then either $\HC$ is not expanding or $\HC$ is not
convex. Assume then that there exists $v\in V(\tbr)$.such that $\hat
C(v) \notin  \{\ddz,d'\}$. Let $U$ the set of such vertices $\{v\in
V(\tbr)\ : \hat
C(v) \notin \{\ddz,d'\}\}$. Then $carrier(U) \cap carrier(d') =
\emptyset$ and also $carrier(U) \cap carrier(\ddz) =
\emptyset$. From $\{\ddz,d'\}$ chose a color whose carrier is
neighboring $carrier(U)$ (if $\ddz,d' \notin \HC(V)$ chose $\ddz$
). Now, let $C'$ be obtained from $\HC$ by coloring the vertices of
$U$ by the color chosen and all the rest retain their color under
$\HC$. Then $cost(C') \le cost(\HC)$ and both conditions of the
observation hold.
endignore 1108}
\end{proof}

The function $REDUCE$ in Subcase 3b is based on the following
observation: Let $C'$ be any
optimal recoloring of $T$ satisfying Observation \ref{obs:wlog}, and let
$s$ be the parent of $\rdz$ in $T$.
Then $C'|_{V(\tbr)}$,
the restriction of the coloring $C'$ to the vertices of $\tbr$,
depends only on whether $carrier(d',C')$ intersects ${V(\tht)}$,
and in this case if it contains the vertex $s$.
Specifically, $C'_{V(\tbr)}$ must be
one of the three colorings of $V(\tbr)$,
$\chig,\cmed$ and $\cmin$, according to the following three scenarios:
\ignore{1127
The function $REDUCE$ in Subcase 3b is based on the following
observation: Let $C'$ be any
optimal recoloring of $T$ satisfying Observation \ref{obs:wlog}, and let
$s$ be the parent of $\rdz$ in $T$.
Then $C'|_{V(\tbr)}$,
the restriction of the coloring $C'$ to the vertices of $\tbr$,
depends only on whether $C'|_{V(\tht)}$ contains a $d'$-block,
and on the location of this $d'$-block.
Specifically, $C'_{V(\tbr)}$ must be
one of the three colorings of $V(\tbr)$,
$\chig,\cmed$ and $\cmin$, according to whether $\tht$ has a
$d'$-block, and in this case whether this block is disjoint from $s$.
The details follow:
endignore 1127}
\begin{enumerate}
\item
$carrier(d',C')\cap V(\tht)\neq\emptyset$ and $s\notin
carrier(d',C')$.
Then it must be the case that $C'$ colors all
the vertices in $V(\tbr)$ by $\ddz$.
This coloring of $\tbr$ is denoted as
$\chig$.
\item
$carrier(d',C')\cap V(\tht)\neq\emptyset$ and $s\in
carrier(d',C')$. Then
$C'|_{\tbr}$ is a coloring of minimal possible cost of $\tbr$
which either equals $\chig$ (i.e. colors all vertices by
$\ddz$), or otherwise colors $\rdz$ by $d'$.
This coloring of $\tbr$ is called $\cmed$.
\item
$carrier(d',C')\cap V(\tht) = \emptyset$. Then $C'|_{\tbr}$
must be an optimal convex recoloring of $\tbr$ by the two colors
$\ddz,d'$.
This coloring of $\tbr$ is called $\cmin$.
\end{enumerate}

We will show soon that the colorings $\chig,\cmed$ and $\cmin$
above can be computed in linear time.
The function $REDUCE$ in Subcase 3b modifies the tree $T$ by replacing
$\tbr$ by a subtree $\tbrz$ with only 2 vertices, $\rdz$ and $\vvz$,
which encodes the three colorings $\chig,\cmed,\cmin$. Specifically,
$REDUCE(T,C,w)=(T',C',w_1)$ where
(see Figure \ref{fig:3-approx-case-3b-reduce}):
\begin{itemize}
\item
$T'$ is obtained from $T$ by replacing the subtree $\tbr$ by the
subtree $\tbrz$ which contains two vertices: a root $\rdz$ with a
single descendant $\vvz$.
\item
$w_1(v)=w(v)$ for each $v\in V(\tht)$.
For $\rdz$ and $\vvz$,
$w_1$ is defined as follows:
$w_1(\rdz)=cost(\cmed)-cost(\cmin)$ and
$w_1(\vvz)=cost(\chig)-cost(\cmin)$. 
\item
$C'(v)=C(v)$
for each
$v\in V(\tht)$;
If $w(\rdz)>0)$ then $C'(\rdz)=\ddz$ and if $w(\vvz)>0$ 
then $C'(\vvz)=d'$.
(If $w_1(u)=0$ for
$u\in\{\rdz,\vvz\}$, then $C'(u)$ is undefined).
\end{itemize}

Figure~\ref{fig:3-approx-case-3b-reduce} illustrates $REDUCE$
for case 3b. In the figure, $\chig$ requires overwriting all
$d'$ vertices and therefore costs $3$, $\cmed$ requires overwriting
one $d_0$ vertex and costs $2$ and $\cmin$ is the optimal coloring for
$\tbr$ with cost $1$. The new subtree $\tbrz$ reflects these weight
with $w_1(r_{d_0})=\cmed-\cmin=1$ and $w_1(v_0)=\chig-\cmin=2$.

\begin{figure}[hbt]
\begin{center}
\input{3-approx-case-3b-reduce_f1.pstex_t}
\end{center}
\caption{$REDUCE$ of case  3b: $\tbr$ is replaced with
$\tbrz$ where $w_1(r_{d_0})=\cmed-\cmin=1$ and $w_1(v_0)=\chig-\cmin=2$.
\label{fig:3-approx-case-3b-reduce}}
\end{figure}


\begin{claim}\label{OPT-T'-le-OPT-T}
$OPT(T',C',w_1) = OPT(T,C,w)-cost(\cmin)$.
\end{claim}
\begin{proof}
We first show that $OPT(T',C',w_1) \leq OPT(T,C,w)-cost(\cmin)$.
Let $C^*$ be an optimal recoloring of $C$ satisfying Observation
\ref{obs:wlog}, and let $X^*=\XX(C^*)$.
By the discussion above,
we may assume that $C^*|_{V(\tbr)}$ has one
of the forms $\chig,\cmed$ or $\cmin$.
Thus, $X^*\cap {V(\tbr)}$ is either $\xhig,\xmed$ or $\xmin$.
We map $C^*$ to a coloring $C'$
of $T'$ as follows: for $v\in\ V(\tht)$, $C'(v)=C^*(v)$. $C'$ on
$\rdz$ and $\vvz$ is defined as follows:
\begin{itemize}
\item
If $C^*|_{V(\tbr)}=\chig$ then $C'(\rdz)=C'(\vvz)=\ddz$, and
$cost(C'|_{V(\tbr})=w_1(\vvz)$;
\item
If $C^*|_{V(\tbr)}=\cmed$ then $C'(\rdz)=C'(\vvz)=d'$,
and $cost(C'|_{V(\tbr})=w_1(\rdz)$;
\item
If $C^*|_{V(\tbr)}=\cmin$ then $C'(\rdz)=\ddz$, $C'(\vvz)=d'$,
and $cost(C'|_{V(\tbr})=0$.
\end{itemize}
Note that in all three cases, $cost(C')=cost(C^*)- cost(\cmin)$.


The proof of the opposite inequality
$OPT(T,C,w)-cost(\cmin)  \leq OPT(T',C',w_1)$
is similar.
\end{proof}

\begin{corollary}
$C^*$  is optimal recoloring of $(T,C,w)$ iff $C'$ is an optimal
recoloring of $(T',C',w_1)$.
\end{corollary}

We now can define the $UPDATE$ function for Subcase 3b: Let
$X'=\aprx(T',C',w_1)$. Then $X'$ is a disjoint union of the sets
$\xpht=X'\cap V(\tht)$ and $\xpbrz=X'\cap V(\tbrz)$. Moreover,
$\xpbrz\in \{\{\rdz\},\{\vvz\},\emptyset\}$.
Then $X\leftarrow UPDATE(X')= \xpht\cup \xpbr$, where $\xpbr$ is
$\xhig$ if $\xpbrz=\{\rdz\}$, is $\xmed$ if $\xpbrz=\{\vvz\}$, and
is $\xmin$ if $\xpbrz=\emptyset$. Note that
$w(X)=w(X')+cost(\cmin)$.
The following inequalities show that if $w_1(X')$ is a
3-approximation to $OPT(T',C',w_1)$, then $w(X)$ is a 3-approximation
to $OPT(T,C,w)$:
\begin{eqnarray*}
w(X)&=& w_1(X')+cost(\cmin)\leq 3OPT(T',C',w_1)+cost(\cmin)\\
&<& 3(OPT(T',C',w_1)+cost(\cmin))=3OPT(T,C,w)
\end{eqnarray*}

\subsection{A Linear Time Algorithm for Subcase 3b}
In subcase 3b we need to compute $\chig$, $\cmed$ and $\cmin$.
The computation of $\chig$ is immediate. $\cmed$ and $\cmin$
 can be computed by the following simple, linear time
algorithm that finds a minimal cost convex recoloring
of a bi-colored tree, under the constraint that
the color of a given vertex $r$
is predetermined to one of the two colors.\\
Let the weighted colored tree $(T,C,w)$ and the vertex $r$ be given,
and let $\{d_1,d_2\}=C(T)$.
For $i\in\{1,2\}$, let $C_i$
the minimal cost convex recoloring 
which sets the color of $r$ to $d_i$ 
(note that
a coloring with minimum cost in $\{C_1,C_2\}$
is an optimal convex recoloring of $(T,C)$).
We illustrate the computation of $C_1$ (the computation of $C_2$ is
similar):\\
%
Compute for every edge $e=(u \rightarrow v)$ a cost defined by
$$
cost(e) =  \displaystyle  w(\{v':v' \in T(v)\text{ and }
C(v')=d_1\}) +
\displaystyle w(\{v': v' \in [T\setminus T(v)] \text{ and }
C(v')=d_2\})
$$
where $T(v)$ is the subtree rooted at $v$.
This can be done by one post order traversal of the tree.
Then, select the edge $e^* =(u_0\rightarrow v_0)$
which minimizes this
cost, and set $C_1(w)= d_2$ for each $w\in T(v_0)$, and $C_1(w)=d_1$
otherwise.

\subsection{Correctness and complexity}
We now summarize the discussion of the previous section
to show that the algorithm terminates and return a
cover $X$ which is a 3-approximation for $(T,C,w)$.

Let $(T=(V,E),C,w)$ be an input to \aprx. if
$V\setminus support(w)$ is a cover then the returned solution is
optimal. Else, in each of the cases,
$REDUCE(T,C,w)$ reduces the input to
$(T',C',w_1)$ such that $|support(w_1)|<|support(w)|$, hence the
algorithm terminates within at most $n=|V|$ iterations. Also, as
detailed in the previous subsections, the
function $UPDATE$ guarantees that that if $X'$ is a 3-approximation
for $(T',C',w_1)$ then $X$ is a 3-approximation to $(T,C,w)$. Thus
after at most $n$ iterations the algorithm provides a 3-approximation
to the original input.

Checking whether Case 1, Case 2, Subcase 3a or Subcase 3b
holds at each stage requires
$O(cn)$ time for each of the cases, and computing the function $REDUCE$
after the relevant case is identified requires linear time in all cases.
Since there are at most $n$ iterations, the overall
complexity is $O(cn^2)$. Thus we have
\begin{theorem}
Algorithm \aprx\  is a polynomial time
3-approximation algorithm for the minimum
convex recoloring problem.
\end{theorem}

\section{Discussion and Future Work}\label{sec:conc}
In this work we showed two approximation algorithms for colored
strings and trees, respectively. The 2-approximation algorithm relies on
the technique of penalizing a colored string and the 3-approximation
algorithm for the tree extends the local ratio technique by allowing
dynamic changes in the underlying graph.

Few interesting research directions which suggest themselves are:
\begin{itemize}
\item
Can our approximation ratios for strings or trees be improved. 
\item
This is a more focused variant of the previous item.
A problem has a {\em polynomial approximation scheme}
\cite{GarJoh79,Hoc97}, or is {\em fully
approximable} \cite{PazMoran81},
if for each $\varepsilon$ it can be
$\varepsilon$-approximated in
$p_{\varepsilon}(n)$ time for some polynomial $p_{\varepsilon}$.
Are the problems of optimal convex recoloring of trees or strings fully
approximable, (or equivalently have
a polynomial approximation scheme)?
\item Alternatively, can any of the variant be shown to be
APX-hard \cite{}?
\item The algorithms presented here apply only to uniform models. The
  {\em non uniform model},
 motivated by weighted maximum parsimony
\cite{san75}, assumes that
the cost of assigning color $d$ to vertex $v$
is given by an arbitrary
nonnegative number $cost(v,d)$ 
(note that, formally, no
initial coloring $C$ is assumed in this cost model).
In this model $cost(C')$ is defined only for a
total recoloring $C'$, and is given by the sum
$\sum_{v\in V}cost(v,C'(v))$. Finding non-trivial approximation
results for this model is challanging.

\end{itemize}
\ignore{end
In this work we tackled the case of introducing a new character to an
existing phylogenetic tree. We considered three variants of the
general case: In the first two variants, the input is different but
the objective function is the same - minimizing the state changes at
the vertices. In the third variant, the objective function is
different. We showed that these three variants are NP-Complete and
devised a general algorithm to solve all of them. We also gave FPT
algorithms for for the first two variants and a $2$-approximation
algorithm for the string case and a  $3$-approximation
algorithm for the tree case under the first variant.\\
In spite of the hardness results, it would be interesting to reduce
the space complexity of our exact algorithms, devise a more efficient
algorithms for the particular cases under the different model variants
and look for a better approximation.
endignore end}

\section{Acknowledgments}
We wish to thank Reuven Bar Yehuda and Mike Steel for many helpful
discussions.

\bibliography{sagi}
\bibliographystyle{alpha}

\clearpage

\end{document}